\documentclass[journal,draftcls,onecolumn,12pt,twoside]{IEEEtranTCOM}
\UseRawInputEncoding
\usepackage{inputenc}
\usepackage{graphicx}
\usepackage{epsfig}
\usepackage{subfigure}
\usepackage{cite}
\usepackage{setspace}
\usepackage{url}
\usepackage{amssymb}
\usepackage{amsmath}
\usepackage{color}
\usepackage{booktabs}
\usepackage{epstopdf}
\usepackage{bigstrut}
\usepackage{bm}
\usepackage{enumerate}
\usepackage{algorithm}
\usepackage{algorithmic}
\usepackage{amsmath}
\usepackage{xcolor}
\begin{document}

% paper title
% can use linebreaks \\ within to get better formatting as desiblack
\title{GAN-based Generator of Adversarial Attack on Intelligent End-to-End Autoencoder-based Communication System}

\author{Jianyuan Chen,Lin Zhang,~\IEEEmembership{Senior Member,~IEEE},Zuwei Chen,~\IEEEmembership{Graduate Student Member,~IEEE},Yawen Chen, Hongcheng Zhuang~\IEEEmembership{Member,~IEEE} \\
	\thanks{
		Manuscript received in May 1, 2025. This work was supported by the National Key
Research and Development Program of China under grant 2021YFA0716600. (\textit{Corresponding author: Lin Zhang and Hongcheng Zhuang, e-mail: isszl@mail.sysu.edu.cn; .})
		%\newline
		Jianyuan Chen and Lin Zhang are with School of Cyber Science and Technology, Sun Yat-sen University, Shenzhen, 518107, China (e-mail:chenjy596@mail2.sysu.edu.cn).
		Zuwei Chen is with the Department of Electrical Engineering, City University of Hong Kong, Hong Kong SAR, China (e-mail: zuweichen2-c@my.cityu.edu.hk).
		Yawen Chen is with School of Systems and Computing, University of New South Wales, Canberra, 2612,  Australia (e-mail: wendy.chen1@unsw.edu.au).
Hongcheng Zhuang is with School of Electronics and Communication Engineering, Sun Yat-sen University, Shenzhen 518033, China. (e-mail:zhuanghch@mail.sysu.edu.cn).
	}
}

\maketitle

%%%%%%%%%%%%%%%%%%%%%%%%%%%%%%%%%%%%%%%%%%%%%%%%%%%%%%%%%%%%%%%%%%%%%%%%%%%%%%%%%%%%%%%%%%%%%%%%%%%%%%%%%%%%%%%%%%%%%%%%%%%%%%%%%%%%%%%%%%
%%%%%%%%%%%%%%%%%%%%%%%%%%%%%%%%%%%%%%%%%%%%%%%%%%%%%%%%%%%%%%%%%%%%%%%%%%%%%%%%%%%%%%%%%%%%%%%%%%%%%%%%%%%%%%%%%%%%%%%%%%%%%%%%%%%%%%%%%%
\begin{abstract}
Deep neural networks have been applied in wireless communications system to intelligently adapt to dynamically changing channel conditions, while the users are still under the threat of the malicious attacks due to the broadcasting property of wireless channels. However, most attack models require the knowledge of the target details, which is difficult to be implemented in real systems. Our objective is to develop an attack model with no requirement for the target information, while enhancing the block error rate. In our design, we propose a novel Generative Adversarial Networks(GANs) based attack architecture, which exploits the property of deep learning models being vulnerable to perturbations induced by dynamically changing channel conditions. In the proposed generator, the attack network is composed of convolution layer, convolution transpose layer and linear layer. Then we present the training strategy and the details of the training algorithm. Subsequently, we propose the validation strategy to evaluate the performance of the generator. Simulations are conducted and the results show that our proposed adversarial attack generator achieve better block error rate attack performance than that of benchmark schemes over Additive White Gaussian Noise (AWGN) channel, Rayleigh channel and High-Speed Railway channel.
\end{abstract}
%%%%%%%%%%%%%%%%%%%%%%%%%%%%%%%%%%%%%%%%%%%%%%%%%%%%%%%%%%%%%%%%%%%%%%%%%%%%%%%%%%%%%%%%%%%%%%%%%%%%%%%%%%%%%%%%%%%%%%%%%%%%%%%%%%%%%%%%%%
%%%%%%%%%%%%%%%%%%%%%%%%%%%%%%%%%%%%%%%%%%%%%%%%%%%%%%%%%%%%%%%%%%%%%%%%%%%%%%%%%%%%%%%%%%%%%%%%%%%%%%%%%%%%%%%%%%%%%%%%%%%%%%%%%%%%%%%%%%
%%%%%%%%%%%%%%%%%%%%%%%%%%%%%%%%%%%%%%%%%%%%%%%%%%%%%%%%%%%%%%%%%%%%%%%%%%%%%%%%%%%%%%%%%%%%%%%%%%%%%%%%%%%%%%%%%%%%%%%%%%%%%%%%%%%%%%%%%%
%%%%%%%%%%%%%%%%%%%%%%%%%%%%%%%%%%%%%%%%%%%%%%%%%%%%%%%%%%%%%%%%%%%%%%%%%%%%%%%%%%%%%%%%%%%%%%%%%%%%%%%%%%%%%%%%%%%%%%%%%%%%%%%%%%%%%%%%%%
\vspace{-0.2cm}
\section{Introduction}
\label{sec:intro}
Mobile computing has revolutionized the way we live and work. Portable computing devices such as smartphones, tablets, and laptops can access information and perform tasks on the go. They enable the people to stay connected, access emails, browse the internet, and use various applications at any time. This flexibility has transformed the workplace, allowing professionals to work remotely and collaborate with colleagues from different locations.

In mobile computing systems, to improve the communication performances including the reliability and the data rate, deep learning has been applied to intelligently adapt to dynamically changing channel conditions. With the aid of powerful representation learning capabilities, deep learning has been used in practical systems such as physical layer communication systems\cite{ch12}\cite{ch14}, resource allocation systems\cite{ch15}\cite{ch16}, and intelligent traffic control systems\cite{ch17}. Besides, the deep neural network has also been utilized to construct the end-to-end autoencoder-based communication system to provide end-to-end optimization of communication performances.

However, the natural broadcasting property of wireless channels brings a big challenge to the security of information transmissions. Mobile devices might suffer from eavesdropping or malicious attacks. Moreover, for communication systems applying the deep learning technology, introducing a small perturbation to the neural network's input can easily lead to erroneous output, thereby affecting the security of the entire communication systems. Especially for high-speed vehicular communications, it is crucial to protect real-time data exchanges, ensure safe vehicle operations, prevent cyberattacks, and maintain traffic safety and efficiency.

Recently, the impact of adversarial attacks has become increasingly prominent, especially for the intelligent deep learning aided wireless communications. To study the attack behavior characteristics, researchers have put efforts to build up attack models. For example, in single-input single-output systems, Sadeghi et al.\cite{ch28} propose an iterative approach to produce a generic perturbation that can spoof the channel decoder independently of its input. Cao et al.\cite{ch25} propose a C\&W-based attack algorithm which is destructive to end to-end communication systems from the perspectives of white-box attacks and black-box attacks.

Besides, inspired by previous generative adversarial networks used to generate synthetic image examples, Shi et al.\cite{ch124} propose a deception attack scheme based on generative adversarial networks. Similarly, Bahramali et al.\cite{ch125} propose an input-independent, undetectable and robust algorithm for adversarial attacks based on generative adversarial networks in both white-box and black-box scenarios. Moreover, the presence of universal adversarial perturbations\cite{ch99} reveals important geometric relationships between the high-dimensional decision boundaries of the classifiers, which further outlines potential security vulnerabilities in the presence of a single direction in the input space.
Furthermore, for Multiple Input Multiple Output (MIMO) systems, Ye et al.\cite{ch122} evaluate different white-box and black-box adversarial attack algorithms for a deep learning-based multi-user Orthogonal Frequency Division Multiplexing (OFDM) detector.

However, these attack methods have two major defects. First, these attack models lack the ability to adapt to the dynamic and highly mobile channel variations of vehicular networks, which degrade the practicality. Second, they require complete knowledge of the target model, hence these methods can hardly be applied in real systems that such information is unavailable.

Our motivation is to propose an intelligent attack model to remove the removes the requirement of knowing the transceiver structure, and exploits the vulnerabilities of deep learning models to enhance the attack performance. Different from existing methods, we propose a novel adversarial perturbation generator to increase the block error rate, which can vice versa improve the resilience against potential attacks.

In this design, we propose a novel Generative Adversarial Networks (GANs) based attack model, which exploits the property that the deep neural networks are sensitive to dynamically changing channel conditions, to enhance the attack performance and the practicality. Notably, our input-agnostic attack method allows the intelligent adversarial perturbation generator to operate independently, which are independent from the input signal. This enables the perturbation to be pre-evaluated and integrated seamlessly into the signal prior to the transmission, thus enhancing the flexibility and the practicality of the proposed research work.

Briefly, the main contributions are listed as follows.
\begin{enumerate}
	\item We design the architecture of the attack model, wherein the GANs based adversarial perturbation generator are serially connected with the deep learning-based autoencoder communication systems. Moreover, we design the training algorithm for the proposed model.
	
	\item We propose a novel mapping rule, which can generate perturbations from input data which has lower dimension than the benchmark schemes. Besides, applying the mapping rule, the perturbation increases the difficulty for detection mechanisms to identify and mitigate the adversarial perturbation.
	
	\item We propose an efficient loss function to accelerate the training of the adversarial perturbation generator, with the aim to improve the performance of the adversarial attack. Moreover, using the proposed loss function, the optimization can be conducted more efficiently, thus the attack capabilities can be further enhanced.
\end{enumerate}

The remainder of this paper is organized as follows. Section II presents the intelligent end-to-end autoencoder communication system design and the proposed attack model. Section III elaborates the training algorithms, and analyzes the complexity. Subsequently, in Section IV, the simulation model is described, and the attack performances are simulated and analyzed over Additive White Gaussian Noise (AWGN), Rayleigh fading and high speed railway channels. Finally, Section V concludes this paper.

%%%%%%%%%%%%%%%%%%%%%%%%%%%%%%%%%%%%%%%%%%%%%%%%%%%%%%%%%%%%%%%%%%%%%%%%%%%%%%%%%%%%%%%%%%%%%%%%%%%%%%%%%%%%%%%%%%%%%%%%%%%%%%%%%%%%%%%%%%
%%%%%%%%%%%%%%%%%%%%%%%%%%%%%%%%%%%%%%%%%%%%%%%%%%%%%%%%%%%%%%%%%%%%%%%%%%%%%%%%%%%%%%%%%%%%%%%%%%%%%%%%%%%%%%%%%%%%%%%%%%%%%%%%%%%%%%%%%%

\vspace{-0.2cm}
\section{System Model}
\label{sec:system-model-and-formulate}
In this section, we present the intelligent autoencoder communication system, then we introduce the channel model and analyze the vulnerability of the neural networks.

\subsection{Intelligent end-to-end autoencoder communication system}

\begin{figure}[htbp]
	\centering
	\includegraphics[width=0.8\linewidth]{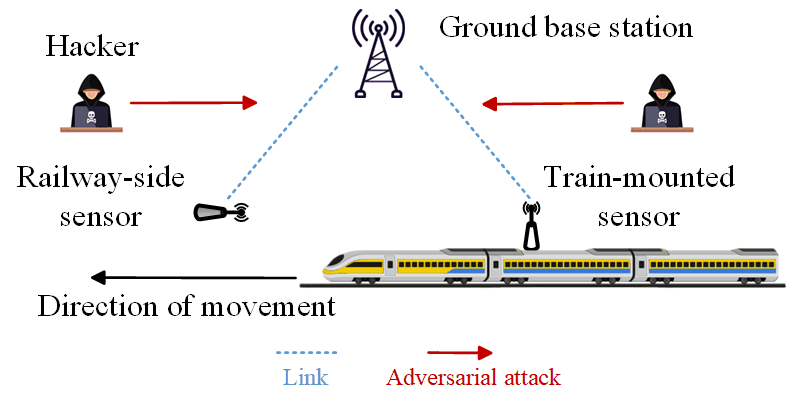}\\
	\vspace{-0.2cm}
	\caption{Application scenario of the intelligent end-to-end autoencoder-based communication system model}
	\label{fig_1}
\end{figure}

Figure \ref{fig_1} illustrates the application scenario of the intelligent end-to-end autoencoder communication system which transmits the information based on the deep learning based transceivers. In this intelligent communication based railway transportation network, sensors collect the data characterizing the operation status. For example, along the railway tracks, multiple Railway Side Sensors (RSs) are strategically placed. These RSs are sensing the surrounding environmental data as comprehensive as possible to avoid missing the key information. Simultaneously, on the running trains, multiple Train-borne Sensors (TSs) are actively engaged in collecting all data reflecting the status of the train. Notably, both the RSs and TSs systematically transmit the sensed data to the Ground Base Station (BS).

Notably, since wireless channels naturally have the broadcasting property, the sensed data, which are the key to maintain the normal operation of the railway system, will be under the risk of being eavesdropped or attacked by malicious users. Hence how to guarantee the security of these data and the data transmission is vital issue for the railway transportation network.

\begin{figure}[htbp]
	\centering
	\includegraphics[width=0.9\linewidth]{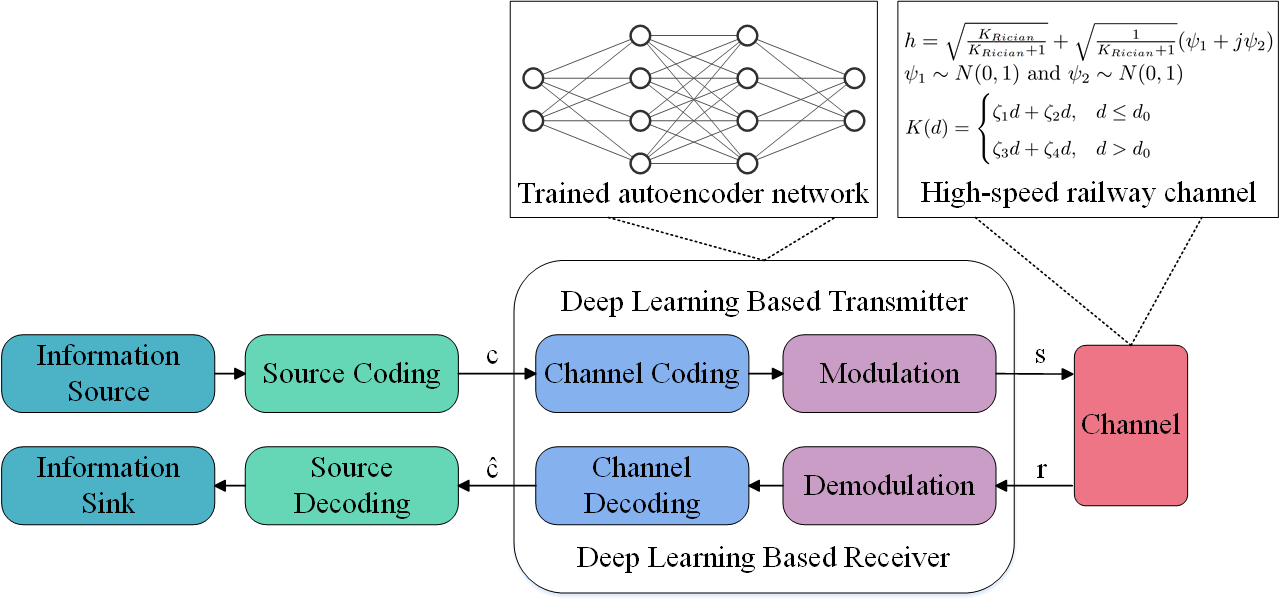}\\
	%\vspace{-0.2cm}
	\caption{Diagram of Data Transmission in High-speed Railway Communication System}
	\label{fig0}
\end{figure}

Figure \ref{fig0} illustrates the data transmission model of a high-speed railway communication system. It shows the process from the Information Source to the Information Sink. In detail, let \(c\) denote the message from source and \(\hat{c}\) represent the estimates obtained from the receiver. The end-to-end DL-based transmission system mainly consists of three key components, which are transmitter, channel, and receiver.

In the transmitter component, the message \(c\) that is intended to be sent undergoes a series of linear and nonlinear operations, through which the message gets mapped into numbers. Thus the transmitter component functions as a mapping denoted by \(F(\cdot)\) to generate the transmitted signal \(s = F(c)\). Then the information-bearing signals are delivered via the channel undergoing the additive noises as well as multiplicative perturbations. Subsequently, the received signal is obtained as:
\begin{equation}
	r = H(s)
\end{equation}
where \(H(\cdot)\) is a function characterizing the channel component, which includes the additive noises represented by \(n\) and the channel responses represented by \(h\).

In the receiver component, the estimates \(\hat{c}\) corresponding to the transmitted message \(c\) can be retrieved from \(r\) after conducting both linear and nonlinear operations. Suppose \(I(\cdot)\)  represents the mapping from \(r\) to \(\hat{c}\), we will have:
\begin{equation}
	\hat{c} = I(r).
\end{equation}

\subsection{Channel model}
In the considered railway communication scenario, the user terminals such as trains have high mobility, we will apply the single-path fast Rayleigh fading channel and the high-speed railway channel with AWGN to simulate the information-bearing channels.

To be explicit, for the AWGN channel, the received signal \(r\) is expressed as \(r = s + n\), where \(s\) denotes the transmitted signal, and \(n\) represents the additive white Gaussian noise.
In this channel model, the noise \(n\) has a probability density function given by \(p(n)=\frac{1}{\sqrt{2\pi\sigma^{2}}}\exp\left(-\frac{n^{2}}{2\sigma^{2}}\right)\), where \(\sigma^{2}\) is the variance of the noise. The mean of the noise is zero.

For the single-path fast Rayleigh fading channel, the received signal \(r\) is represented as \(r = h\odot s + n\), where the channel coefficient \(h\) is expressed as \(h=\psi_{1}+j\psi_{2}\), \(\psi_{1}\) and \(\psi_{2}\) are independent and identically distributed Gaussian random vectors with a mean of \(0\) and a variance of \(1\). The normalized amplitude \(|h|\) of the channel coefficient \(h\) follows a Rayleigh distribution. Considering the mobility in high-speed vehicle communication, it is assumed that the channel coefficient changes in each time slot to reflect the fast fading characteristic. Due to the large variations in space and speed during vehicle information transmission, the channel coefficients in \(h\) are mutually independent, and the single-coefficient exponential correlation matrix is an identity matrix with a correlation coefficient of \(0\).

For the high-speed railway channel, existing research works indicate that the Rician distribution can better describe the channel conditions. The channel parameter \(h\) consists of a constant component due to the line-of-sight (LOS) propagation and a Rayleigh component due to the non-line-of-sight (NLOS) propagation, which is defined as
$h=\sqrt{\frac{K_{Rician}}{K_{Rician}+1}}+\sqrt{\frac{1}{K_{Rician}+1}}(\psi_{1}+j\psi_{2})$. The normalized amplitude \(|h|\) of the channel parameter \(h\) follows a Rician distribution. The Rician \(K\) factor can be modeled as a two-slope linear function. Let \(K(d)\) be the Rician \(K\) factor as a function of distance \(d\), and the function is given by:
\[K(d)=\begin{cases}
	\zeta_{1}d+\zeta_{2}d, & d\leq d_{0}\\
	\zeta_{3}d+\zeta_{4}d, & d > d_{0}
\end{cases}\]
where $\text{d}$ and $\text{d}_{0}$ denote the transmitter-to-receiver distance and the break-point distance respectively.

In this paper, the suburban high-speed railway scenario is considered, thus $\zeta_{1}=-0.027 \text{dB}$, $\zeta_{2}=8.48 \text{dB}$, $\zeta_{3}=-0.0023 \text{dB}$, $\zeta_{4}=4.024 \text{dB}$, $\text{d}_{0} = 200 \text{m}$, and the mean value of Rician $K$-factor $K_{\text{Rician}}=2.83  \text{dB}$ \cite{ch100} is adopted.

\subsection{Vulnerability analysis of neural networks}
\label{sec21}

\begin{figure}[htbp]
	\centering
	\includegraphics[width=1\linewidth]{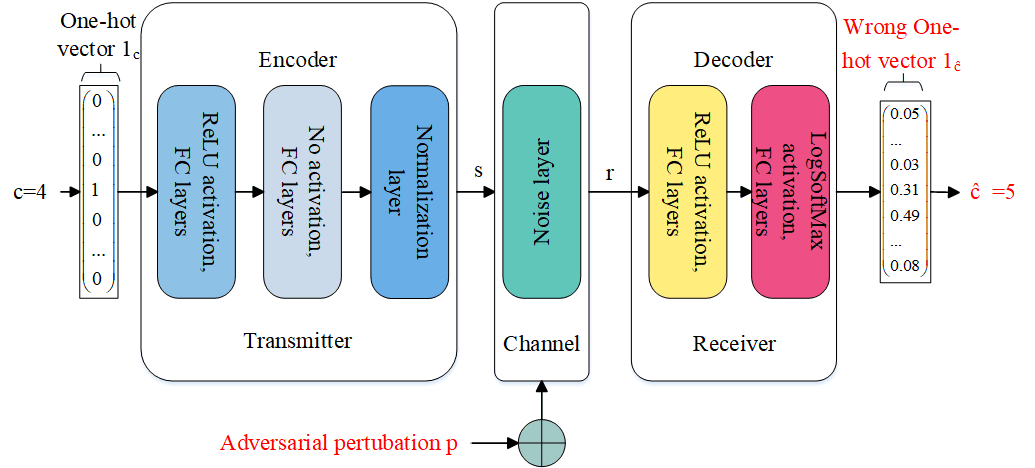}\\
	\vspace{-0.2cm}
	\caption{Attack imposed on intelligent receivers}
	\label{fig1}
\end{figure}

As shown in Fig. \ref{fig1}, an attacker can exploit the broadcast nature of the channel to transmit the adversarial perturbation $p$ into the channel. The received signal $r_{\text{adv}} = H(s) + p$ undergoing the perturbation will cause the intelligent decoder to produce incorrect decoding results.

Adversarial perturbation is a carefully crafted vector that is imperceptible to the neural network but highly sensitive. The goal of the adversarial attack is to find an perturbation $p$ for a given sample $r$ and construct an adversarial example:
\begin{equation}
	\label{opt1}
	\min ||r_{\text{adv}} - r|| < \rho
\end{equation}
and
\begin{equation}
	\label{opt2}
	I(r_{\text{adv}}) \neq I(r)
\end{equation}
where $|| \cdot ||$ represents the chosen distance metric, $\rho$ is the maximum imperceptible perturbation based on that metric, \(I(\cdot)\)  represents the NN decoder. It can be seen that the neural networks are sensitive to perturbations, which can be utilized to design adversarial attacks to be presented as follows.

\section{Adversarial perturbation generator}
\label{sec:Optimal Resource Allocation}
In this section, we will propose the adversarial perturbation generator, and the training algorithms. Then we will present the details of the loss function design, and the deployment of the attack generator.

\subsection{Training framework and algorithm design}

%修改图片
\begin{figure}[htbp]
	\centering
	\includegraphics[width=1\linewidth]{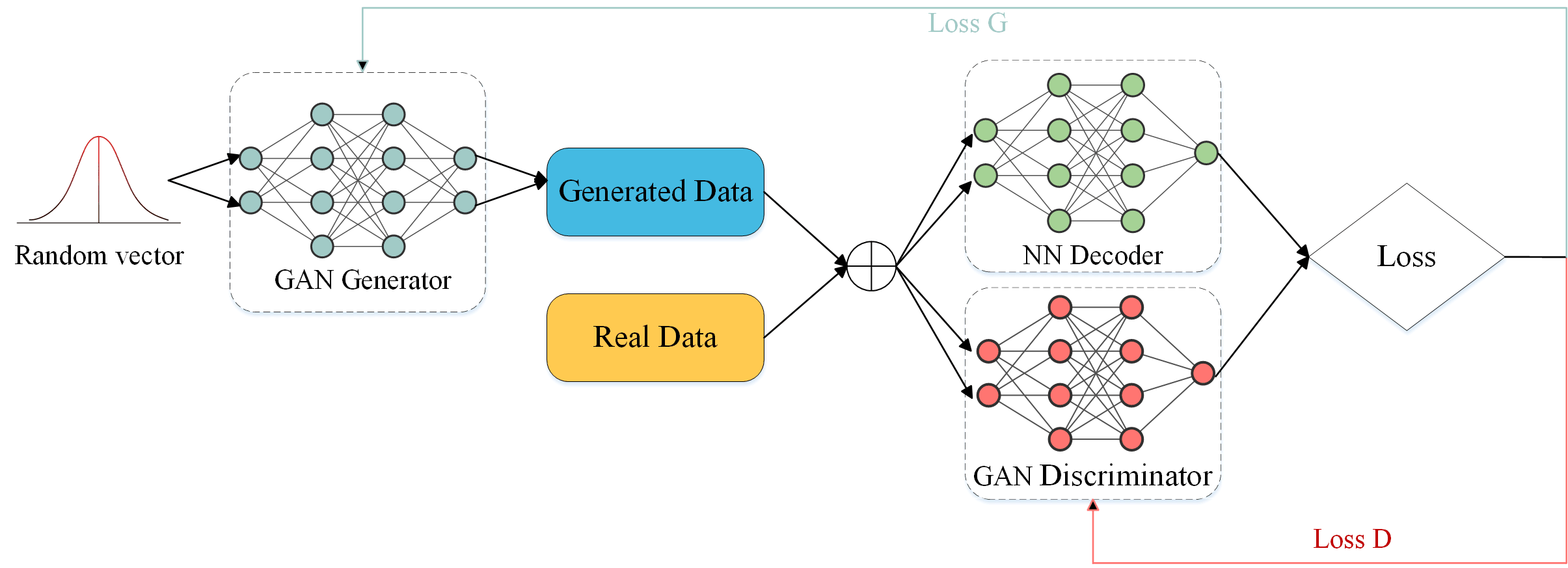}\\
	\vspace{-0.2cm}
	\caption{Training framework of adeversarial attack generator}
	\label{fig2}
\end{figure}

We design the training framework of the intelligent adversarial perturbation generator based on the AdvGAN framework\cite{ch26}, as shown in Fig. \ref{fig2}. Suppose \(G(\cdot)\) represents the GAN generator mapping from vector \(m\) to adversarial perturbation \(p\) and \(D(\cdot)\) represents the GAN discriminator mapping from vector to probability. In our design, firstly, to ensure that the perturbations generated by the generator are difficult to detect, the adversarial sample is designed as $r + G(m)$ instead of the original $G(r)$. In this structure, the generator receives a low-dimensional random Gaussian noise vector $m$ and maps it to the original sample space, generating a realistic adversarial sample $r_{\text{adv}} = r + G(m)$. During the training, the discriminator attempts to accurately distinguish between the samples $r_{\text{adv}}$ generated by the generator and real data samples $r$. This competitive training process encourages the generator to continually improve the quality of the generated samples, producing adversarial perturbations to initiate attacks on the targets.

\subsection{Structure of adversarial perturbation generator}
By exploiting the vulnerability of the neural network being sensitive to dynamically changing channel conditions, we propose to compose an adversarial perturbation attack generator.

\begin{figure}[htbp]
	\centering
	\includegraphics[width=0.5\linewidth]{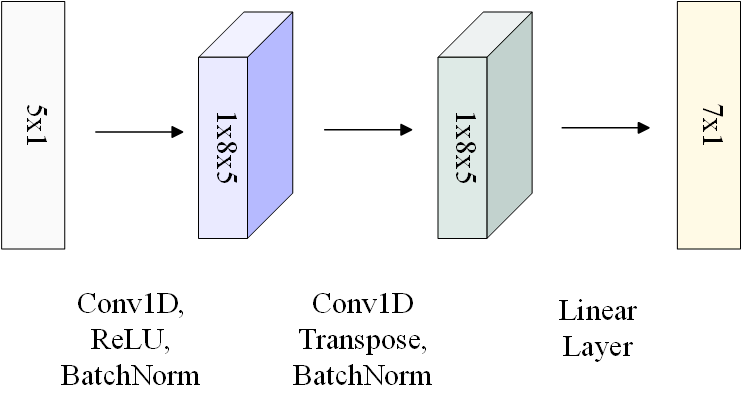}\\
	\vspace{-0.2cm}
	\caption{Structure of the generator}
	\label{fig8}
\end{figure}

\begin{figure}[htbp]
	\centering
	\includegraphics[width=0.5\linewidth]{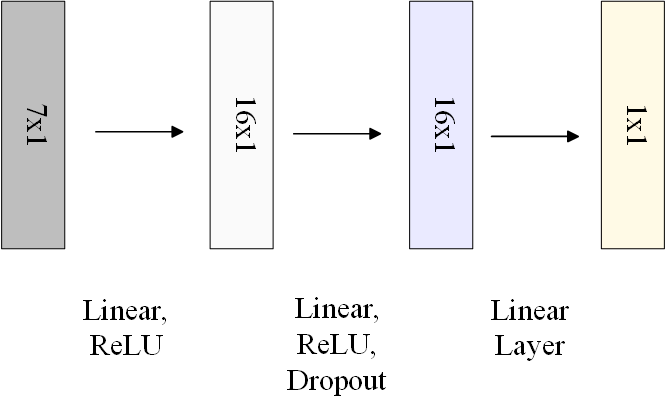}\\
	\vspace{-0.2cm}
	\caption{Structure of the discriminator}
	\label{fig9}
\end{figure}

Considering that Generative Adversarial Networks (GANs) can generate high-quality, diverse samples without the need to explicitly define the probability density function of the distribution \cite{ch24}, we propose to construct an intelligent adversarial perturbation generator based on GANs in the end-to-end autoencoder system. Fig. \ref{fig8} and Fig. \ref{fig9} show our network structure. Based on the structure of intelligent NN decoder shown in Fig. \ref{fig1}, we design the GAN generator and discriminator to achieve better attack performance.

\subsection{Loss function design}
Although the generator trained under the AdvGAN framework can produce high-quality and diverse samples, its training process is unstable and difficult to converge. To address this issue, we design the loss functions for the generator and discriminator based on the Wasserstein distance instead of JS divergence, which can solve the problem of the mode collapse \cite{arjovsky2017wassersteingan}.

To be more explicit, the design principle of the loss function is based on the concept of the Lipschitz continuity. That is, for a continuous function \(f: \mathbb{R}^n \rightarrow \mathbb{R}^m\), it is required that there exists a constant \(K\) such that for any two elements \(x\) and \(y\) in the domain, the following holds:
\begin{equation}
	\|f(x) - f(y)\| \leq K \|x - y\|
\end{equation}
where \(\|\cdot\|\) represents the norm of a vector. Here \(K\) is called the Lipschitz constant of the function \(f\). The intuitive explanation of Lipschitz continuity is that the rate of change of the function \(f\) is limited over the entire domain, and its rate of change will not exceed \(K\) times the rate of change of the input variable.

The Wasserstein distance is defined as the supremum of the expected differences of all Lipschitz-continuous functions \(f\), and this distance quantifies the minimum cost of transforming one distribution into another, which can be expressed as follows:
\begin{equation}
	W(P_r, P_g) = \sup_{\|f\|_L \leq K} \mathbb{E}_{r \sim P_r}[f(r)] - \mathbb{E}_{r \sim P_g}[f(r)]
\end{equation}
where \(P_{r}\) is the distribution of the real data, \(P_{g}\) is the distribution of the output data of the generator, $\mathbb{E}[\cdot]$ represents the evaluation of the numerical expectation and \(f\) is a Lipschitz-continuous function.

Then we exploit the fitting ability of neural networks to design a discriminator \(f\) with parameters \(w\). During the training process, all parameters \(w_{i}\) of the neural network \(f_{\theta}\) are constrained not to exceed the range \(\left[-c,c\right]\). At this time, the partial derivative \(\frac{\partial f_{w}}{\partial r}\) with respect to the input sample \(r\) is bounded, thereby meeting the Lipschitz continuity condition. When the discriminator is used to fit the function \(f\), the Wasserstein distance can be approximated as:
\begin{equation}
	W(P_r, P_g) \approx \max_{\|f_{w}\|_L \leq K} \mathbb{E}_{r \sim P_r}[f_{w}(r)] - \mathbb{E}_{r \sim P_g}[f_{w}(r)]
\end{equation}

The goal of the discriminator is to estimate the Wasserstein distance as accurately as possible, that is:
\begin{align}
	\min_{\|f_{w}\|_L \leq K} \mathbb{E}_{r \sim P_g}[f_{w}(r)] - \mathbb{E}_{r \sim P_r}[f_{w}(r)]\\
	\min_{\|f_{w}\|_L \leq K} \mathbb{E}_{r \sim P_g}[D(r)] - \mathbb{E}_{r \sim P_r}[D(r)]\\
	\min_{\|f_{w}\|_L \leq K} \mathbb{E}_{r \sim P_r}[D(r + G(m))] - \mathbb{E}_{r \sim P_r}[D(r)]\\
	\min_{\|f_{w}\|_L \leq K} \mathbb{E}_{r \sim P_r}[D(r + G(m)) - D(r)]
\end{align}
Here \(L_{D} = D(r + G(m)) - D(r)\) is applied to attain higher accruacy.

The goal of the generator is to reduce the Wasserstein distance as smaller as possible to generate samples with higher precision. Considering that \(\mathbb{E}_{x \sim P_r}[f_{w}(x)]\) is independent of the generator, the optimization goal is set as:
\begin{align}
	\min_{\|f_{w}\|_L \leq K} -\mathbb{E}_{r \sim P_g}[f_{w}(r)]\\
	\min_{\|f_{w}\|_L \leq K} -\mathbb{E}_{r \sim P_g}[D(r)]\\
	\min_{\|f_{w}\|_L \leq K} -\mathbb{E}_{r \sim P_r}[D(r + G(m))]
\end{align}

As described in Section \ref{sec21}, the design goal of adversarial attacks is to achieve imperceptibility and misclassification. In this research work, we further optimize the generator's loss function by dividing it into two parts: $L_{\text{G1}}$ and $L_{\text{G2}}$. The imperceptibility goal constrained by Equation \ref{opt1} is equivalent to minimizing the Wasserstein distance, thus we propose that $L_{G1} = -D(r + G(m))$.

Furthermore, the misclassification goal constrained by Equation \ref{opt2} is equivalent to maximizing the difference between the distribution of adversarial samples generated by the generator and the true label distribution. Considering that the cross-entropy can measure the difference based on the probability distribution of the data, we adopt the negative of the cross-entropy function as the loss function, i.e., $L_{\text{G2}} = \displaystyle\sum_{i} c_i \log(I(r_i + G(m_i)))$, where $c_i$ is the label of $r_i$, \(I(\cdot)\)  represents the NN decoder.

By weighting the losses of the two tasks, $L_{\text{G1}}$ and $L_{\text{G2}}$, the generator's loss function is formulated as follows:
\begin{equation}
	\label{criteron_G}
	L_{\text{G}} = -\lambda D(r + G(m)) + (1-\lambda) \displaystyle\sum_{i} c_i \log(I(r_i + G(m_i)))
\end{equation}

\subsection{Training and validation algorithm}
At the training stage of adversarial attack generator shown in Fig. \ref{fig2}, we use two optimizers $optimizer_{\text{D}}$ and $optimizer_{\text{G}}$ to update the parameters of GAN discriminator and generator. The discriminator’s parameters are updated after each backpropagation step, while the generator’s parameters are updated only after a specified number of iterations of the discriminator.
%补充表述

At the validation stage of adversarial attack generator, we firstly generate the low-dimensional random Gaussian variable $z$ for each test sample. Then the perturbations are generated using the pre-trained generator model, and added to the test samples to create adversarial samples. Finally, the intelligent decoder is used to estimate the blocks, then the block error rate can be evaluated.

The training and validation algorithms designed for the intelligent adversarial perturbation generator are as follows:

\begin{algorithm}[h]
	\caption{Training Algorithm for Intelligent Adversarial perturbation Generator}
	\label{gan_train}
	\begin{algorithmic}[1]
		\STATE \textbf{Input:} Learning rate $\alpha$, clipping parameter $c$, batch size $m$, number of discriminator iterations per generator iteration $n\_critic$, latent space noise dimension $latent\_dim$
		\FOR{each training iteration}
		\FOR{each batch sample $real\_batch$ and its corresponding index $i$}
		\STATE \# Clear discriminator gradients
		\STATE optimizer\_D.zero\_grad()
		\STATE \# Generate latent space noise
		\STATE $m = \text{randn}(0, 1, (\text{real\_batch.shape[0]}, \text{latent\_dim}))$
		\STATE \# Compute fake batch
		\STATE $fake\_batch = real\_batch + generator(m)$
		\STATE \# Compute discriminator loss
		\STATE $loss\_D = \text{mean}(critrtion\_D(real\_batch, fake\_batch))$
		\STATE \# Backpropagate to compute gradients
		\STATE $loss\_D.backward()$
		\STATE \# Update discriminator parameters
		\STATE optimizer\_D.step()
		\FOR{each discriminator parameter $p$}
		\STATE \# Clamp parameters
		\STATE $p.clamp(-c,c)$
		\ENDFOR
		\IF{$i$ is a multiple of $n\_critic$}
		\STATE \# Clear generator gradients
		\STATE optimizer\_G.zero\_grad()
		\STATE \# Compute fake batch
		\STATE $fake\_batch = real\_batch + generator(m)$
		\STATE \# Compute generator loss
		\STATE $loss\_G = \text{mean}(critrtion\_G(real\_batch, fake\_batch))$
		\STATE \# Backpropagate to compute gradients
		\STATE $loss\_G.backward()$
		\STATE \# Update generator parameters
		\STATE optimizer\_G.step()
		\ENDIF
		\ENDFOR
		\ENDFOR
	\end{algorithmic}
\end{algorithm}

\begin{algorithm}[h]
	\caption{Validation Algorithm for Intelligent Adversarial perturbation Generator}
	\label{gan_valid}
	\begin{algorithmic}[1]
		\STATE \textbf{Input:} Codeword test sample TEsample, corresponding information bit original value TEoriginal, latent space noise dimension latent\_dim, trained adversarial perturbation generator model Generator
		\FOR{each test sample $\text{TEsample}$}
		\STATE \# Generate latent space noise
		\STATE $m = \text{randn}(0, 1, (\text{TEsample.shape[0]}, \text{latent\_dim}))$
		\STATE \# Compute perturbation
		\STATE $\text{perturbation} = \text{Generator}(m)$
		\STATE \# Get bit estimate
		\STATE $\text{TEestimated} = \text{argmax}(\text{Decoder\_NN}(\text{TEsample} + \text{perturbation}))$
		\STATE \# Compute number of erroneous blocks
		\STATE $\text{error\_block} = \text{error\_block} + (\text{TEestimated} \neq \text{TEoriginal}).\text{sum()}$
		\ENDFOR
		\STATE \# Compute block error rate
		\STATE $\text{BLER} = \text{error\_block} / \text{total\_block}$
	\end{algorithmic}
\end{algorithm}

\subsection{Deployment of attack generator}

\begin{figure}[htbp]
	\centering
	\includegraphics[width=1\linewidth]{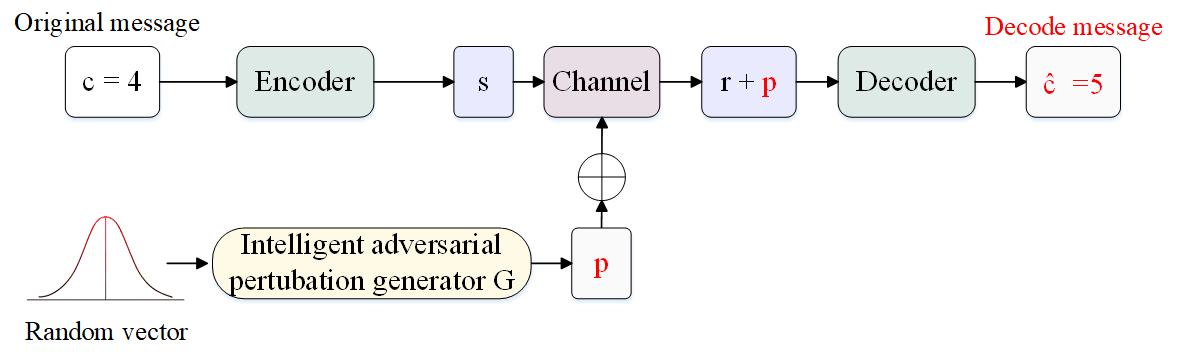}\\
	\vspace{-0.2cm}
	\caption{The intelligent autoencoder-based communication system under adeversarial attacks.}
	\label{fig3}
\end{figure}

Figure \ref{fig3} presents the structure of the autoencoder-based communication system incorporating the intelligent adversarial perturbation generator. As shown in the figure, the intelligent adversarial perturbation generator creates a low-dimensional random Gaussian variable, which is then mapped to an adversarial perturbation and added to the channel, causing the intelligent decoder to produce incorrect decoding results. As a result, the reliability of the autoencoder-based communication system degrades, and legitimate receivers can not retrieve the information reliably, the attack succeeds.

To briefly sum up, the advantages of the improved intelligent adversarial perturbation generator and the training framework in the autoencoder-based communication system compared to the traditional AdvGAN framework are summarized as follows:

1. The input to the generator does not rely on the input signal, allowing the perturbation to be pre-evaluated and added to the signal in advance. Besides, the dimensionality of the generator’s input is lower than that of the original sample, thus the perturbation can be generated from lower-dimensional variables, thus the probability of being detected is lowered.

2. Since the proposed attack model (Equation (\ref{criteron_G})) operates as a gray-box attack, it requires access to critical system components—such as information source bits, and use the input/output of the decoder to train an effective Generator. Malicious attackers, however, typically lack such privileged access to the original data, rendering the training algorithm inapplicable. This design effectively limits the attack's feasibility to authorized testing scenarios, preventing adversaries from exploiting the technology to disrupt communication systems.

\section{Simulation Results}
\label{sec:simu-results}
In this section, we build up the simulation model of the proposed design, and present the details of the training parameter settings.

\subsection{Experimental Environment}
The development environment for the simulation experiments is as follows. The processor model is Intel Core i7 8700; the memory is 32GB; the graphics card is NVIDIA GeForce RTX 1080 Ti; the operating system is Windows 10; the programming languages used are Matlab and Python; and the software platforms are Matlab R2022a and Pycharm.

\subsection{Parameter settings}
In this simulation experiment, autoencoder-based communication system is first trained based on the structure shown in Fig. \ref{fig1}. In the simulation, we set the code length \( n = 7 \) and the number of information bits \( k = 4 \), with the neural network parameters of the intelligent encoder and decoder set as shown in Table \ref{tab31} and Table \ref{tab32}.
\begin{table}[h]
	\centering
	\caption{Structure of Encoder\_NN}
	\label{tab31}
	\resizebox{0.75\textwidth}{!}{  % Resize to 50% width
		\begin{tabular}{ccc}
			\toprule[2pt]
			Layer & Activation Function & Output Dimension \\
			\midrule[2pt]
			Input Layer & None & 16 \\
			\hline
			Fully Connected Layer 1 & ReLU & 16 \\
			\hline
			Fully Connected Layer 2 & None & 7 \\
			\hline
			Batch Normalization Layer & None & 7 \\
			\bottomrule[2pt]
		\end{tabular}
	}
\end{table}

\begin{table}[h]
	\centering
	\caption{Structure of Decoder\_NN}
	\label{tab32}
	\resizebox{0.75\textwidth}{!}{  % Resize to 50% width
		\begin{tabular}{ccc}  % Add vertical lines
			\toprule[2pt]
			Layer & Activation Function & Output Dimension \\
			\midrule[2pt]
			Input Layer & None & 7 \\
			\hline
			Fully Connected Layer 1 & ReLU & 16 \\
			\hline
			Fully Connected Layer 2 & LogSoftmax & 16 \\
			\bottomrule[2pt]
		\end{tabular}
	}
\end{table}

Algorithm \ref{gan_train} is used to train the intelligent perturbation generator, with the dropout probability of the discriminator set to 0.2. The data set consists of labeled samples from the trained intelligent encoder, totaling 100,000 samples. The optimizers for both the generator and discriminator are set to RMSprop, with the generator and the discriminator learning rate set to 0.0005. The clipping parameter \( c \) is set to 0.1. The mini-batch stochastic gradient descent algorithm is used, with the mini-batch size \( batch\_size \) set to 32. The number of discriminator iterations per generator iteration \( n\_critic \) is also set to 5, and the latent space dimension of the noise \( latent\_dim \) is set to 5.

As mentioned in Section \ref{sec21}, we propose to utilize the adversarial perturbations to introduce imperceptible noise, with the aim to block the information recovery of the intelligent decoder. When the power of the perturbation is equal to or lower than the noise level, the perturbation is considered imperceptible. Therefore, the following simulations are conducted under the condition of $PNR<0dB$.

Next, our proposed intelligent adversarial perturbation generator is simulated. The performance of the intelligent adversarial perturbation generator is evaluated by simulating the change in BLER of the autoencoder-based communication system as a function of the perturbation-to-noise ratio (PNR). PNR measures the ratio between the power of the perturbation and the noise, which is expressed as:

\begin{equation} PNR_{[dB]} = 10 \log\left( \frac{P_{\text{perturbation}}}{P_{\text{noise}}} \right) \end{equation}

where $P_{\text{perturbation}}$ is the power of the perturbation, and $P_{\text{noise}}$ is the power of the noise.

\subsection{Block error rate performance analysis}

\begin{figure}[htbp]
	\centering
	\includegraphics[width=0.8\linewidth]{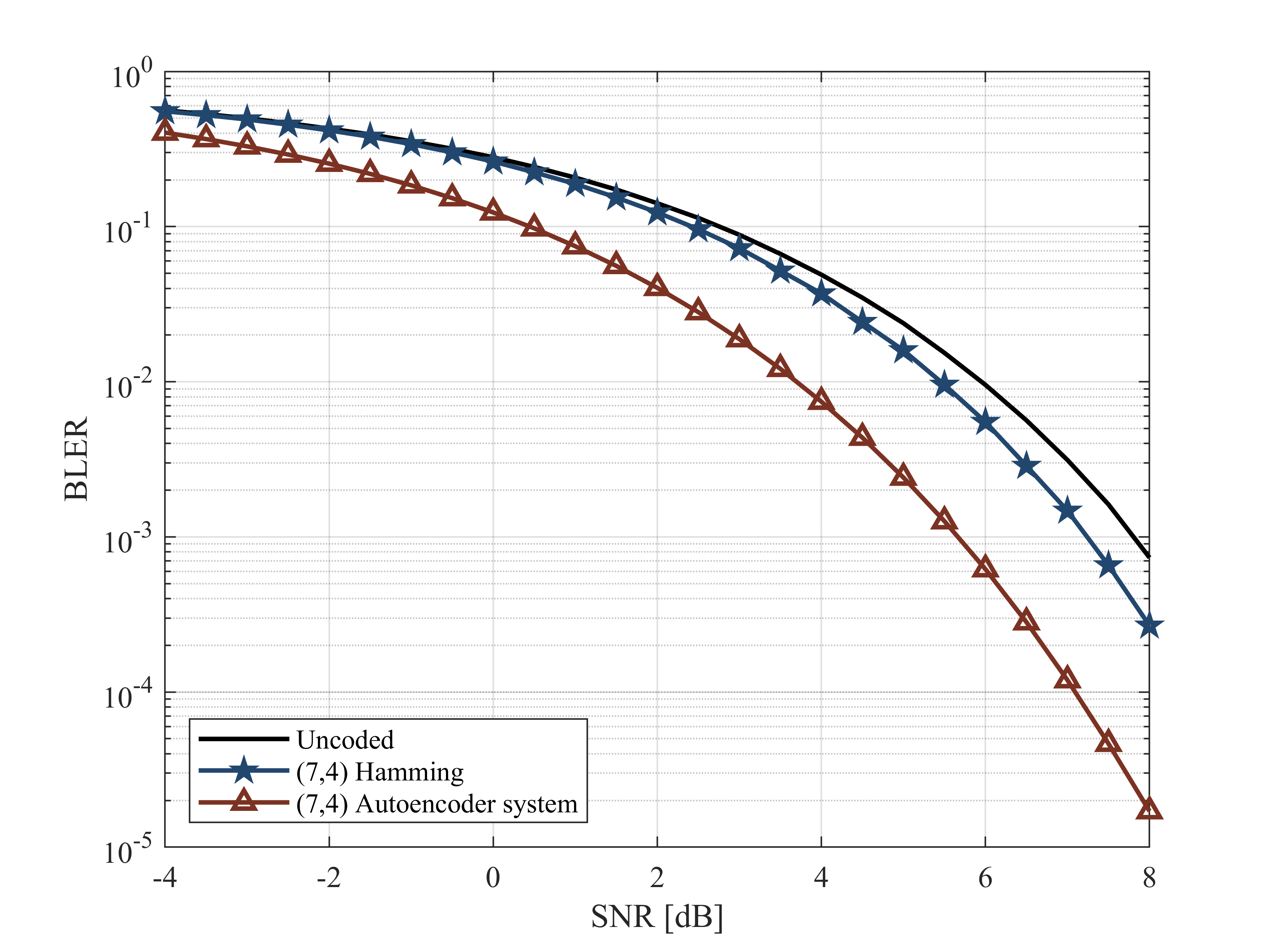}\\
	%\vspace{-0.2cm}
	\caption{BLER performance of the autoencoder-based communication system over AWGN channel}
	\label{fig4}
\end{figure}

\begin{figure}[htbp]
	\centering
	\includegraphics[width=0.8\linewidth]{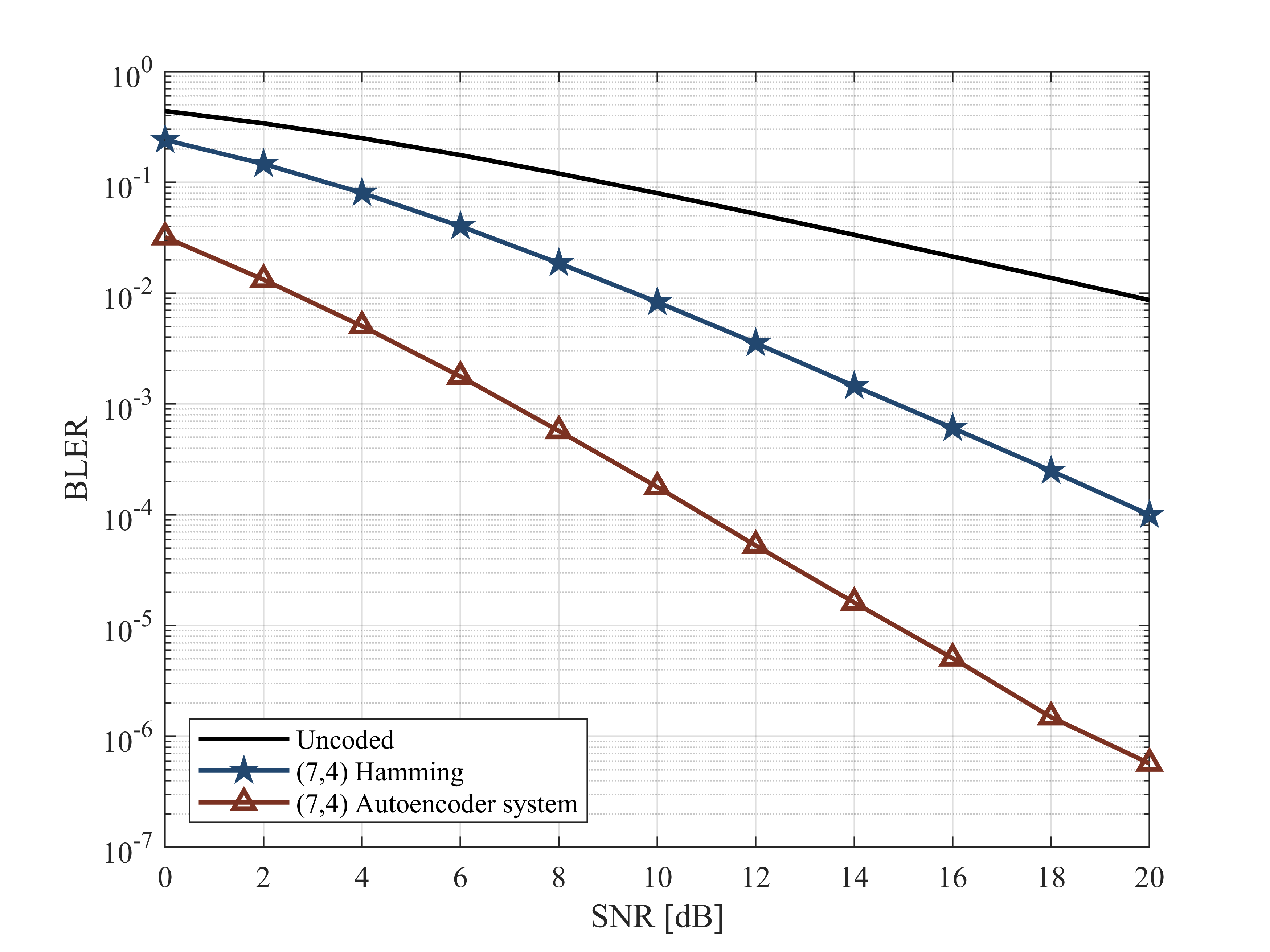}\\
	%\vspace{-0.2cm}
	\caption{BLER performance of the autoencoder-based communication system over Rayleigh channel}
	\label{fig10}
\end{figure}
\begin{figure}[htbp]
	\centering
	\includegraphics[width=0.8\linewidth]{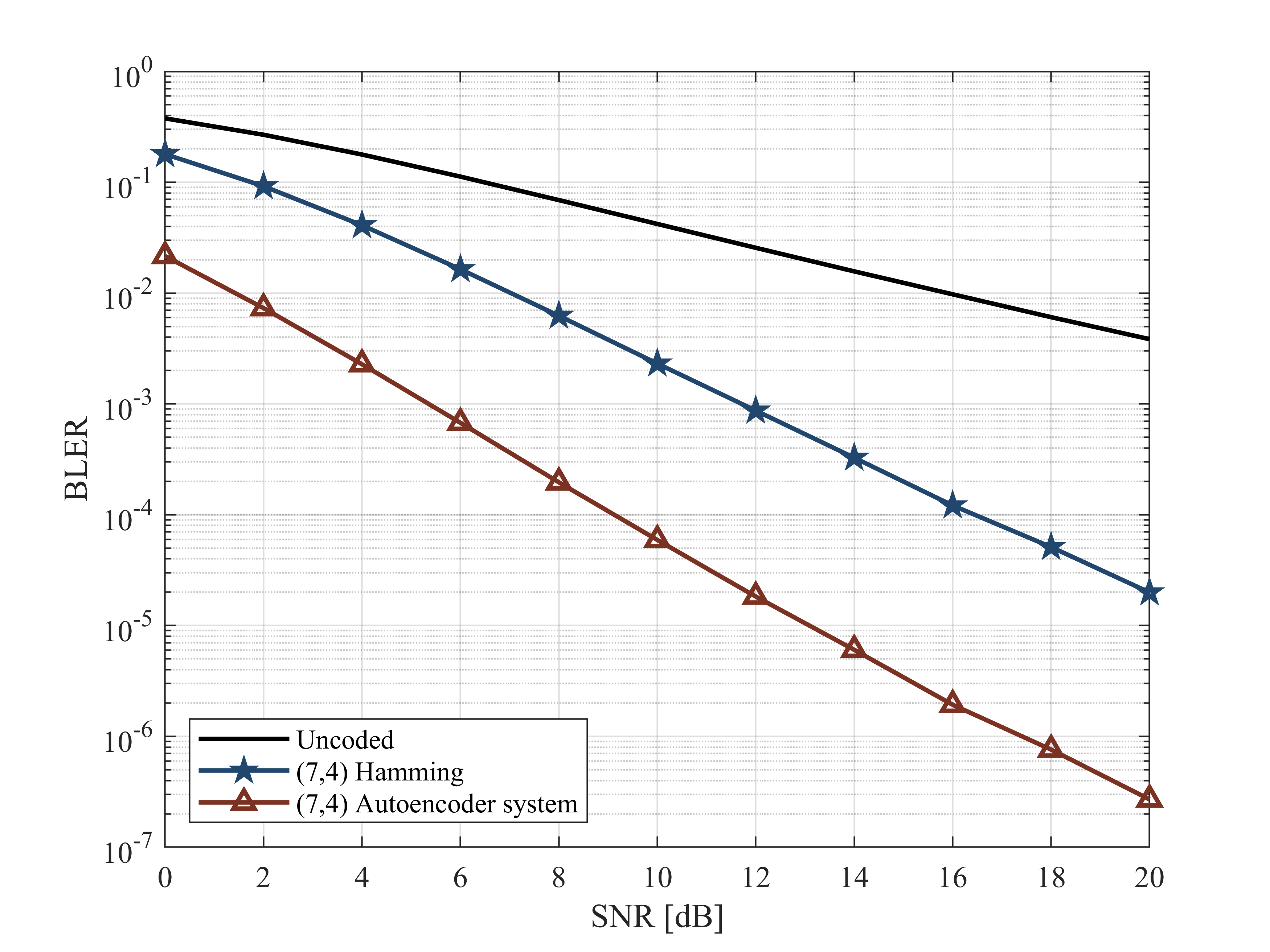}\\
	%\vspace{-0.2cm}
	\caption{BLER performance of the autoencoder-based communication system over High-Speed Railway channel}
	\label{fig11}
\end{figure}

We first analyze the block error rate (BLER) performance of the (7, 4) autoencoder system. As shown in Fig. \ref{fig4}, Fig. \ref{fig10} and Fig. \ref{fig11}, compared with traditional communication system using the binary phase shift keying (BPSK) modulation with hard-decision decoding of the (7, 4) Hamming code, as well as systems using only BPSK modulation, the end-to-end autoencoder-based communication system can achieve better BLER performance.

\subsection{Attack capability analysis and comparisons}

\begin{figure}[htbp]
	\centering
	\includegraphics[width=0.8\linewidth]{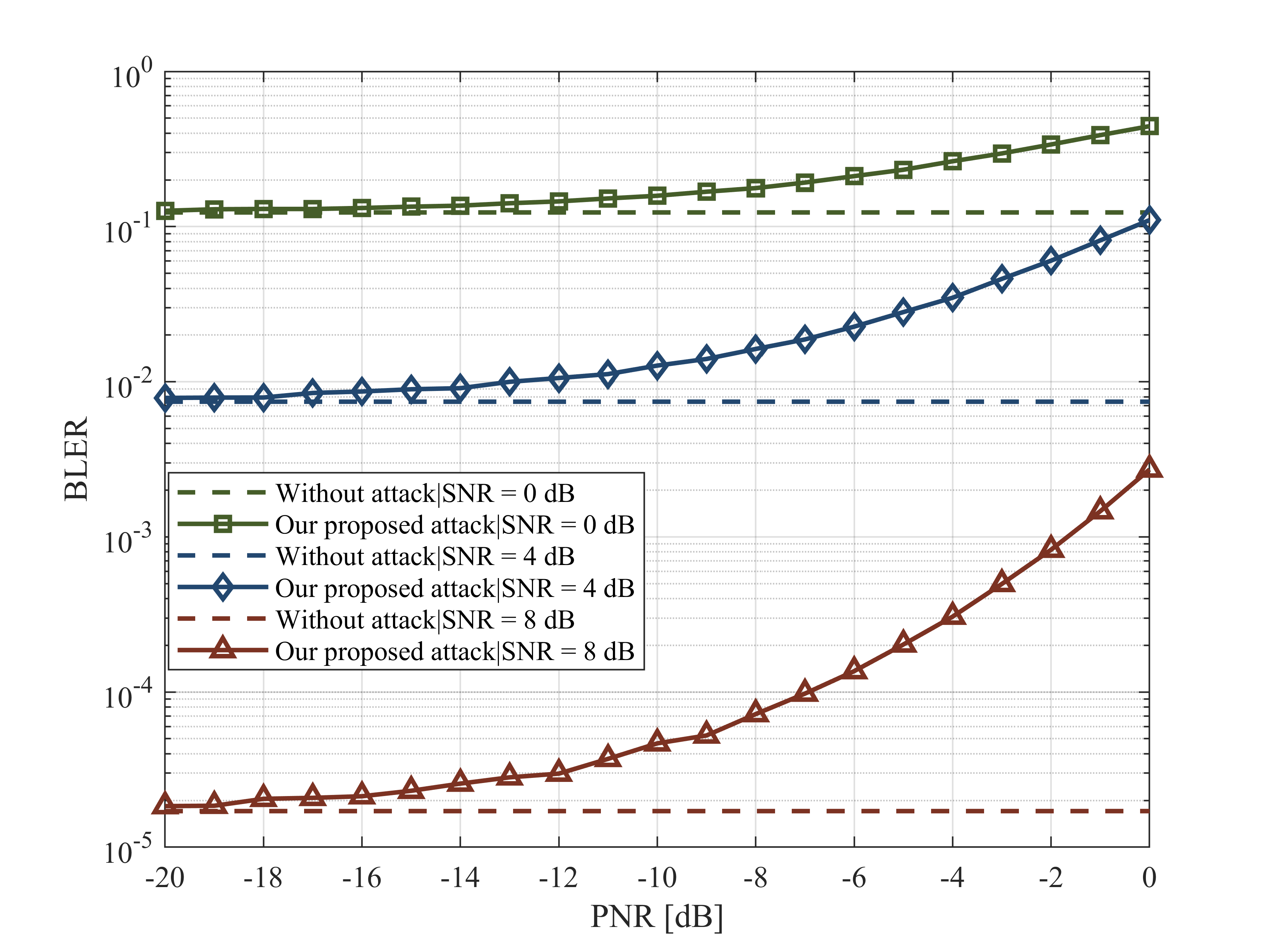}\\
	\vspace{-0.2cm}
	\caption{BLER vs. PNR at different SNR for the system deployed with our proposed attack}
	\label{fig5}
\end{figure}

Figure \ref{fig5} shows the relationship between the system's BLER and PNR when applying the intelligent adversarial perturbation generator under SNR values of 0, 4, and 8 dB, as well as when there is no perturbation. Without adversarial perturbation, the BLER of the autoencoder-based communication system depends on the SNR. We can observe that when the SNR becomes larger, the BLER will be lower, thus better reliability performance can be attained.

Notably, when the system is subjected to adversarial perturbations generated by the intelligent adversarial perturbation generator, the BLER will significantly increase in both high and low SNR scenarios. Under the condition of PNR = 0 dB,  the BLER decreases by 5.6 dB at SNR = 0 dB, by 11.7 dB at SNR = 4 dB, and by 22 dB at SNR = 8 dB. For signals of the same strength, a higher SNR indicates a lower noise level, which means that the machine learning model is more sensitive to the features of the signal. As a result, even small perturbations may lead to a reduction in the model's accuracy. Thus, our proposed intelligent adversarial perturbation generator has a greater impact on the BLER at 8 dB SNR than at 0 dB SNR, indicating that the algorithm will induce a larger BLER degradation in the autoencoder-based communication system, thereby achieving higher attack capability.

\begin{figure}[htbp]
	\centering
	\includegraphics[width=0.8\linewidth]{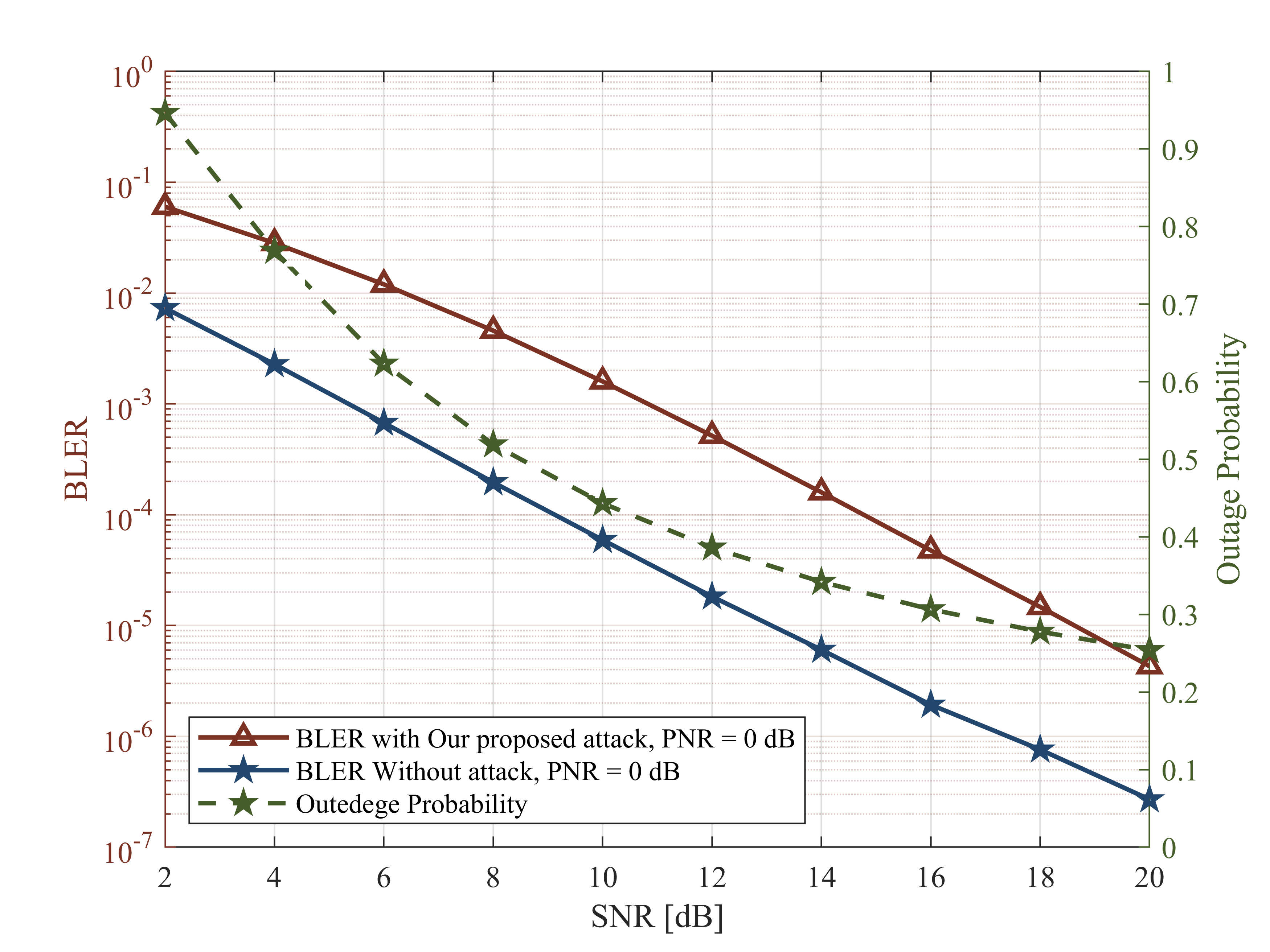}\\
	\vspace{-0.2cm}
	\caption{BLER and Outage Probability vs. SNR with and without attacks}
	\label{fig14}
\end{figure}

Besides, Fig. \ref{fig14} shows the relevant curves of the system BLER and outage probability under the conditions of communication rate of 50Mbps and the bandwidth of 18MHz based on measurement parameters in the Beijing to Tianjin high-speed railway long-term evolution network\cite{ch100}. The abscissa represents the signal-to-noise ratio (SNR) ranging from 2 dB to 20 dB. It can be observed that as the SNR increases, all three curves show a downward trend, and the BLER when being attacked is higher than that without being attacked throughout the entire SNR range. Moreover, the system outage probability is very high at low SNR. In addition, we can notice that when the attack is imposed on the system, the data can not be recovered reliably, and the communication is effectively blocked.

\begin{figure}[htbp]
	\centering
	\includegraphics[width=0.8\linewidth]{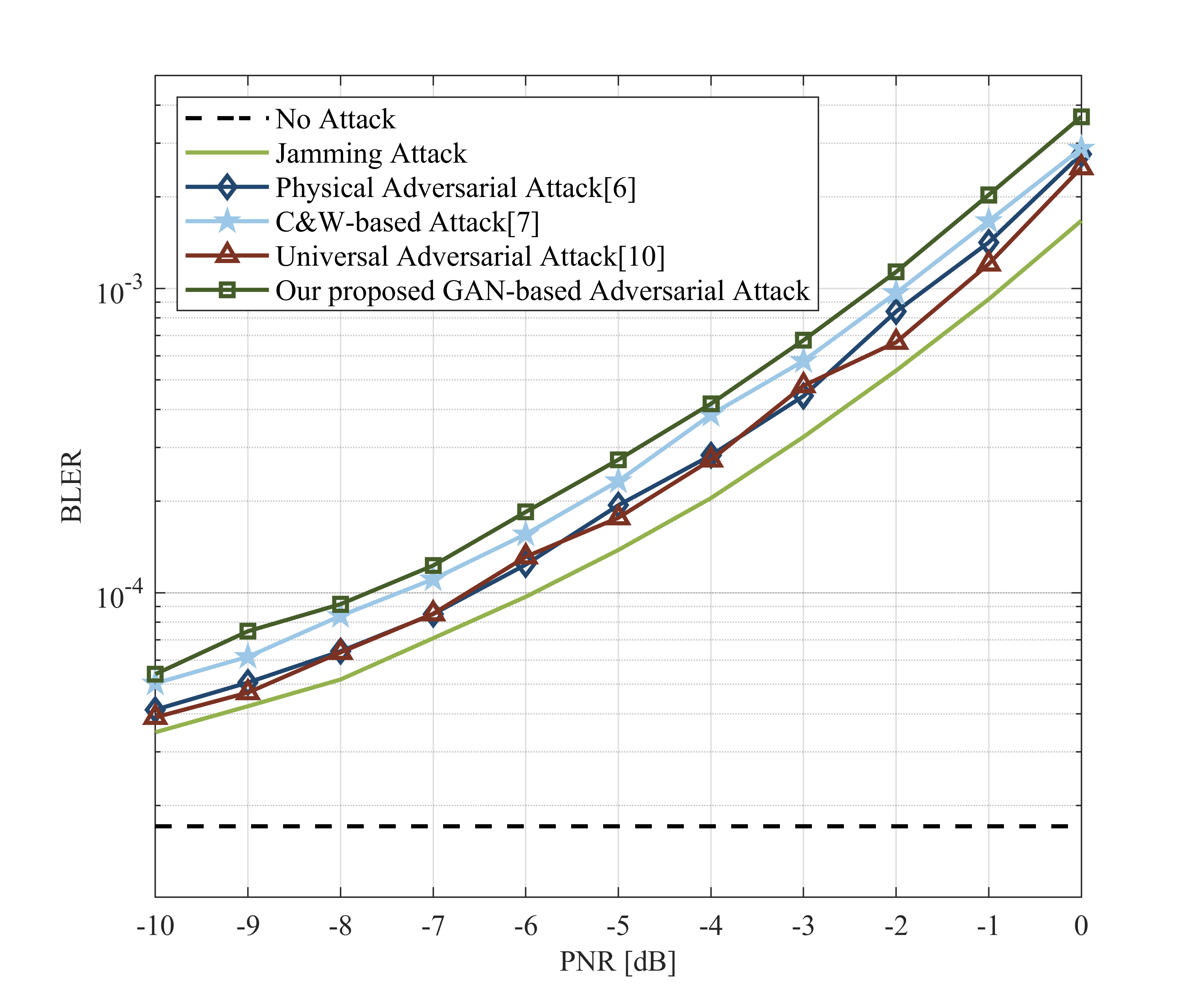}\\
	%\vspace{-0.2cm}
	\caption{Attack performance comparisons in terms of BLER over AWGN channel}
	\label{fig6}
\end{figure}	
\begin{figure}[htbp]
	\centering
	\includegraphics[width=0.8\linewidth]{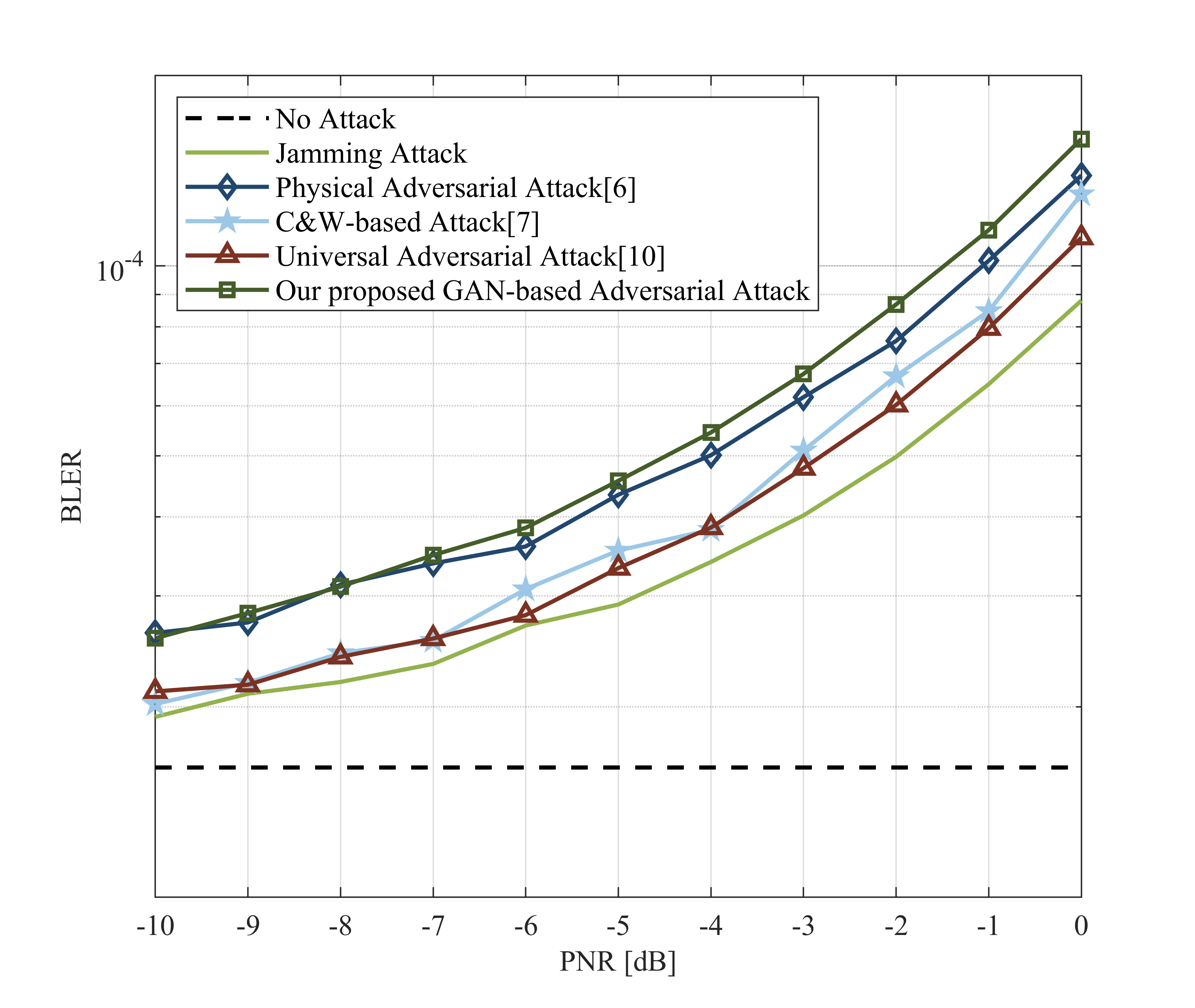}\\
	%\vspace{-0.2cm}
	\caption{Attack performance comparisons in terms of BLER over Rayleigh channel}
	\label{fig12}
\end{figure}	
\begin{figure}[htbp]
	\centering
	\includegraphics[width=0.8\linewidth]{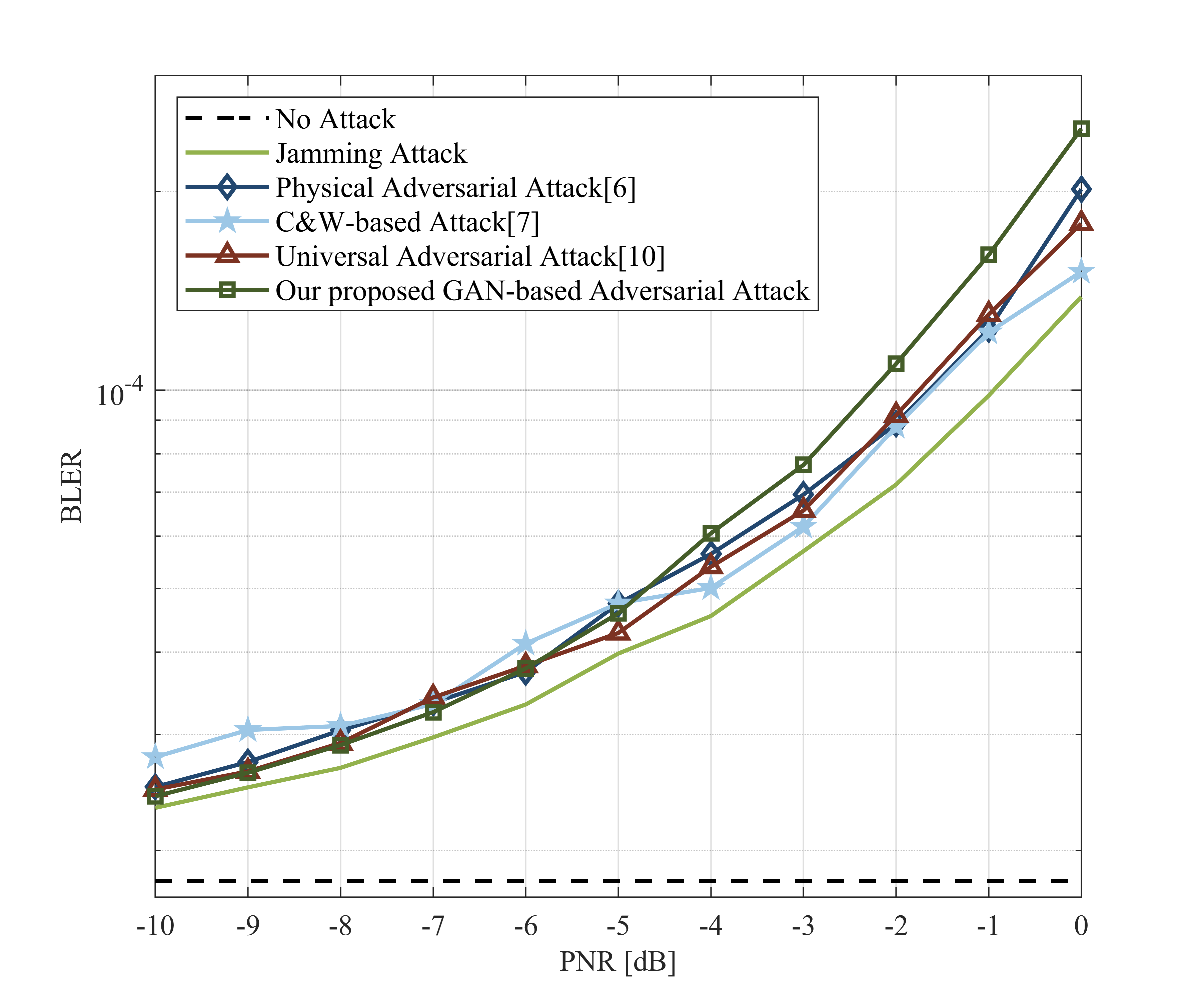}\\
	%\vspace{-0.2cm}
	\caption{Attack performance comparisons in terms of BLER over High-Speed Railway channel}
	\label{fig13}
\end{figure}	

Furthermore, we compare the attack capabilities of the intelligent perturbation generation algorithm with other perturbation generation algorithms. The comparison includes input-agnostic adversarial attack techniques mentioned in paper \cite{ch28} \cite{ch25} and \cite{ch99}, as well as jamming attack which is a simple Gaussian perturbation attack. Fig. \ref{fig6}, \ref{fig12} and \ref{fig13} show the block error rate (BLER) of the system as a function of PNR for different attack methods when SNR = 8 dB over AWGN channel, Rayleigh channel and High-Speed Railway channel.

Across different channel conditions, our proposed GAN-based adversarial attack demonstrates superior attack performance in terms of larger BLER compared to existing methods. To be explicit, when PNR = 0 dB over the AWGN channel, applying our attack model, the BLER increases 119.7\%, which is significantly higher than the BLER of the system undergoing the Physical Adversarial Attack (+65.5\%), C\&W-based Attack (+72.7\%), and Universal Adversarial Attack (+49.4\%).

It is worth pointing out that expect for the AWGN channel, in different application scenarios such as Rayleigh channel and High-Speed Railway channel, the proposed scheme can also attain higher attack performance. More explicitly, in the case of the Rayleigh channel, our method raises the BLER by 80.3\%, outperforming the Physical Adversarial Attack(+57.8\%), C\&W-based Attack(+47.4\%), and Universal Adversarial Attack(+25.8\%). Similarly, in the High-Speed Railway scenario, with our approach, the BLER increases 79.9\%, while the BLER of the system applying the Physical Adversarial Attack, Universal Adversarial Attack, and C\&W-based Attack is respectively +45.6\%, +29.1\% and +9.2\%, which exhibit weaker attack performance.

The reason that better attack performances can be achieved with our design is that the loos function is based on the Wasserstein distance instead of JS divergence, thus the problem of the mode collapse is mitigated, which outperforms the adversarial attack algorithm proposed in \cite{ch28} \cite{ch25} and \cite{ch99}. In addition, we can also notice that the Gaussian perturbation attack, which generates input-agnostic perturbations, poses the least threat to the autoencoder-based communication system. These results consistently highlight that our GAN-based method achieves the strongest adversarial attack over different wireless channels.

\subsection{Covertness analysis}

\begin{figure}[htbp]
	\centering
	\includegraphics[width=0.8\linewidth]{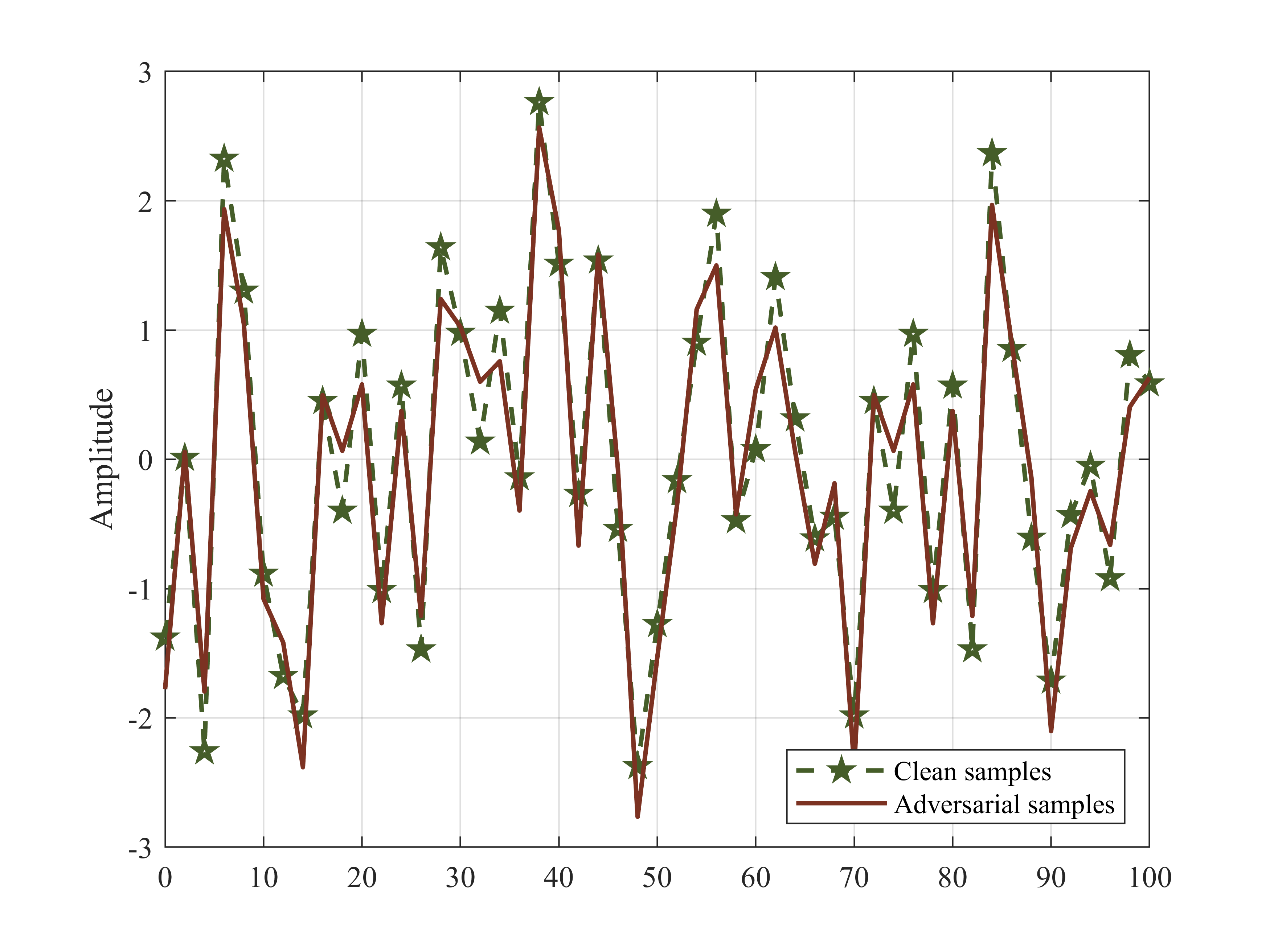}\\
	\vspace{-0.2cm}
	\caption{Time domain waveforms before and after our proposed attack}
	\label{fig7}
\end{figure}

Moreover, we analyze the covertness of the proposed design by comparing the adversarial and clean sample waveforms provided by the intelligent adversarial perturbation generator. Fig. \ref{fig7} compares the time-domain waveforms of adversarial and clean samples when SNR = 8 dB and PNR = 0 dB.

As can be seen from the figure, the waveforms of the clean and adversarial samples almost overlap, with no significant differences in amplitude, frequency, or phase. Therefore, it is difficult to identify or detect which waveform is imposed on the perturbations generated by our proposed intelligent adversarial perturbation generator. Namely, the covertness can be guaranteed, and the intelligent decoder can not realize that the attack has been launched. Thus the block error rate of the autoencoder-based communication system can be enlarged without being recognized, which further increase the probability of successful attacks.

\subsection{Complexity analysis}
At the training stage, the input samples are latent space noise $z$ consisting of variables with length 5 following Gaussian distributions. Since the expected outputs are the fake images with the length of 7, the size of clean batch samples and the fake batch samples is (32, 7). At the deployment stage, we select physical adversarial attack proposed in \cite{ch28} and universarial adversarial perturbation attack proposed in\cite{ch99} as the benchmark.

With the parameter settings presented in TABLE \ref{tab31} and TABLE \ref{tab32}, the computational complexity in terms of the number of trainable parameters and floating point operations per second(FLOPs) of networks in proposed training frameworks is given in TABLE \ref{tab4}. It can be observed that the deep learning aided intelligent encoder and decoder require fewer parameters and FLOPS than the GAN generator and GAN discriminator.

\begin{table}[h]
	\centering
	\caption{COMPLEXITY ANALYSIS OF NETWORKS IN PROPOSED TRAINING FRAMEWORK}
	\label{tab4}
	\resizebox{0.62\textwidth}{!}{  % Resize to 50% width
		\begin{tabular}{ccc}
			\toprule[2pt]
			& \# Trainable params & \# FLOPs \\
			\midrule[2pt]
			GAN Generator & 543 & 50.18K \\
			\hline
			GAN Discriminator & 417 & 12.29K \\
			\hline
			Intelligent Encoder & 133 & 4.48K \\
			\hline
			Intelligent Decoder & 400 & 11.78K \\
			\bottomrule[2pt]
		\end{tabular}
	}
\end{table}

\begin{table}[h]
	\centering
	\caption{Time Consumption Comparisons}
	\label{tab1}
	\resizebox{0.7\textwidth}{!}{  % Resize to 50% width
		\begin{tabular}{cc}  % Add vertical lines on both sides
			\toprule[2pt]
			\textbf{Scheme} & \textbf{Average Time per Sample (ms)} \\
			\midrule[2pt]
			Universal Adversarial Attack[10]  & 308.28 \\
			\hline
			C\&W-based Attack[7]  & 229.68 \\
			\hline
			Physical Adversarial Attack[6]  & 30.14 \\
			\hline
			Our Proposed Attack & 8.07 \\
			\bottomrule[2pt]
	\end{tabular}}
\end{table}
TABLE \ref{tab1} presents and compares the complexity of the proposed design and benchmark schemes. It can be observed from the table that our design achieves higher efficiency, with which adversarial examples can be generated in approximately 8 milliseconds, significantly reducing the sample generation time. By contrast, the method proposed in \cite{ch28} \cite{ch25} and \cite{ch99} take too long time to generate adversarial samples, which potentially allowing the perturbed signal to be correctly decoded before the attack takes effect. Thanks to the short time consumed to generate adversarial examples, higher probability of successful attack will be attained. Hence our design can achieve higher efficiency and better attack capability.

\section{Conclusion}
\label{sec:Conclusion}
In this paper, we propose an efficient intelligent adversarial perturbation generator with higher attack capability. We design a training framework including structure of GAN generator and GAN discriminator constituted by Wasserstein distance, and propose the loss function to improve the convergence and stability of the neural network. Then we analyze the complexity and investigate the attack performance of the proposed design over AWGN channel, Rayleigh fading channel and high speed railway channel. The simulation results show that for the autoencoder-based communication system over AWGN channel, when using the intelligent adversarial perturbation generator with the same perturbation power, the block error rate of the autoencoder-based communication system is higher than benchmark schemes with lower time consumption. Moreover, the perturbation is less detectable, and can be preprocessed to launch the attack as soon as possible, which further enhances the probability of successful attacks. Therefore, for autoencoder-based communication systems, the proposed intelligent perturbation generation algorithm demonstrates stronger attach capabilities compared to other adversarial attack algorithms in terms of block error rate and the efficiency as well as the covertness.

%%%%%%%%%%%%%%%%%%%%%%%%%%%%%%%%%%%%%%%%%%%%%%%%%%%%%%%%%%%%%%%%%%%%%%%%%%%%%%%%%%%%%%%%%%%%%%%%%%%%%%%%%%%%%%%%%%%%%%%%%%%%%%%%%%%%%%%%%%
%%%%%%%%%%%%%%%%%%%%%%%%%%%%%%%%%%%%%%%%%%%%%%%%%%%%%%%%%%%%%%%%%%%%%%%%%%%%%%%%%%%%%%%%%%%%%%%%%%%%%%%%%%%%%%%%%%%%%%%%%%%%%%%%%%%%%%%%%%
%\vspace{-0.2cm}
%%%%%%%%%%%%%%%%%%%%%%%%%%%%%%%%%%%%%%%%%%%%%%%%%%%%%%%%%%%%%%%%%%%%%%%%%%%%%%%%%%%%%%%%%%%%%%%%%%%%%%%%%%%%%%%%%%%%%%%%%%%%%%%%%%%%%%%%%%
%%%%%%%%%%%%%%%%%%%%%%%%%%%%%%%%%%%%%%%%%%%%%%%%%%%%%%%%%%%%%%%%%%%%%%%%%%%%%%%%%%%%%%%%%%%%%%%%%%%%%%%%%%%%%%%%%%%%%%%%%%%%%%%%%%%%%%%%%%
%\bibliographystyle{ieeetr}
\bibliographystyle{IEEEtran}
\bibliography{references_black}

\end{document}